\documentclass[10pt,aps,prd,superscriptaddress,showpacs,twocolumn]{revtex4-1}
\usepackage{graphicx}
\usepackage[caption=false]{subfig}
\usepackage{epstopdf}
\usepackage{amsmath}
\usepackage{amsfonts} 
\usepackage{amssymb}
\usepackage{latexsym}
\usepackage{hyperref}
\usepackage[english]{babel}
\usepackage[utf8]{inputenc}
\usepackage[colorinlistoftodos]{todonotes}
\usepackage{xcolor}
\usepackage{slashed}
\usepackage{bm}

\begin{document}
\title{ $\mathcal{N}=2$ Supersymmetry  with Central Charge: a twofold implementation}


\author{L.P.R. Ospedal}\email{leoopr@cbpf.br}
\affiliation{Centro Brasileiro de Pesquisas F\'{i}sicas (CBPF), Rua Dr. Xavier Sigaud 150, Urca, Rio de Janeiro, Brazil, CEP 22290-180}

\author{R.C. Terin} \email{rodrigoterin3003@gmail.com}
\affiliation{Universidade do Estado do Rio de Janeiro (UERJ),
Instituto de F\'{i}sica, Departamento de F\'{i}sica Te\'{o}rica,
Rua S\~ao Francisco Xavier 524, {Maracan\~{a}}, Rio de Janeiro, Brazil, CEP 20550-013}
\affiliation{Sorbonne Universit\'{e}, CNRS, Laboratoire de Physique Th\'{e}orique de la Mati\`{e}re Condens\'{e}e, \\ LPTMC, F-75005 Paris, France}

\begin{abstract}

In this work, we analyze an extended $\mathcal{N}=2$ supersymmetry with central charge and develop its superspace  formulation under two distinct viewpoints. Initially, in the context of classical mecha-nics, we discuss the introduction of deformed supersymmetric derivatives and their consequence on the  deformation of  one-dimensional  non-linear sigma model. After that, considering a field-theoretical framework, we present an  implementation of this superalgebra in two dimensions, such that one of the coordinates is related to the central charge. As an application, in this two-dimensional scenario, we consider  topological (bosonic) configurations of a special self-coupled matter model and present a non-trivial fermionic solution.

\end{abstract}


\maketitle


\section{Introduction} \label{Sec_Intro}
\indent 

Back to $1975$, the paper by Haag, Lopuszanski and Sohnius \cite{Haag_Lopuszanski_Sohnius} established the most general supersymmetric algebra in four dimensions, which respects the Poincar\'e invariance and requirements of $S-$matrix. In addition to the Poincar\'e operators and usual Majorana supercharges,  it is possible to include new operators, known as central charges, which have all vanishing commutation relations. 

Essentially, the investigations associated with central charges in field theories can be divided in three parts. In the first one, the main subject was the classification of the multiplets  related to massless and massive  representations. In this part, the papers by Salam and Strathdee \cite{Salam_Strathdee} initially analyzed the case without central charges and a series of works clarified the general one  (e.g., see the review \cite{Sohnius_1985} and references therein). These works have shown that, in the presence of central charges, some states are avoided in the massive representation and, in particular situations, central charges may be connected to internal symmetries. 

The second part corresponds to  investigations  at the classical level. One of the remarkable contributions was the work by Witten and Olive \cite{Witten_Olive}, where the authors considered some specific models in two- and four-dimensions with topological configurations and obtained a connection between central charges and topological numbers.  After that, other situations involving topological defects  have been explored, such as complex projective space $CP^{n-1}$ and others
non-linear sigma-models  \cite{Aoyama}$-$\cite{Gracey}. For a more detailed discussion of central charges and topological defects,  we indicate ref. \cite{Hlousek_Spector_1992}.

Other interesting classical point of view  showed up after the dimensional reduction of supersymmetric models. Some works have discussed that central charges can be seen as an inheritance of  dimensional reduction
\cite{Ferrara_Savoy_Zumino}$-$\cite{Gates_1984}. In particular cases, it is related to the momentum operator of the extra-dimension \cite{Alvarez-Gaume_Freedman}. We highlight an exotic situation in ref. \cite{Aoyama_Zizzi}, in which the dimensional reduction of super-Yang-Mills
model in four dimensions with $O(N)$ symmetry group leads to Georgi-Glashow-like model in three dimensions,
where the central charge is associated with the electrical charge of the abelian
subgroup, i.e., other non-topological explanation to central charge. We  also indicate some supersymmetric models with central charges in higher-dimensional theories and brane-world scenarios \cite{Ferrara_zumino_79}$-$\cite{Gorsky_Shifman_2000}. 

The third part concerns to the quantum aspects. Specifically, by means of quantum effects, the central charge has appeared as a quantum anomaly in  superalgebra. From this perspective, many situations have been analyzed, such as non-linear sigma models, kink, monopoles, domain walls and vortices \cite{Losev_Shifman_PRD}$-$\cite{Shifman_Vainshtein_Zwicky}.

In the context of supersymmetric mechanics with central charge,   some investigations of field-theoretical models may correspond to (quantum) mechanical systems, namely, in the study of superconducting cosmic strings, localized fermions on domain walls, gapped and superconducting graphene \cite{Oikonomou}. Moreover, in some well-known mechanical situations, such as Coulomb, Aharonov–Bohm–Coulomb  and Aharonov–Casher systems, it is possible to accommodate  extended supersymmetries with central charge \cite{Niederle_Nikitin_Z}, which explain the degeneracy of the energy spectra. In the presence of an external electromagnetic background,  the Poincar\'e algebra  leads to a residual symmetry algebra with a central charge, deformed translations and Lorentz generators \cite{Karat,Carrion_Rojas_Toppan}.

At this point, it is worthy to emphasize that we are using the term “central charge" based on the generators with all vanishing commutation relations. However, there are situations in which new bosonic generators have been added to  superalgebra with some non-vanishing commutators. For example, we address to the cases of weak supersymmetry \cite{Smilga_2004} and other generalizations known as centrally extensions involving the $su(2|1)$ and $su(2|2)$ algebras \cite{Ivanov_Sidorov,Ivanov_Lechtenfeld_Sidorov}. 
Finally, we highlight some results in supersymmetric mechanics in which the introduction of central charges can be related to duality and mirror symmetries \cite{Ivanov_Krivonos_Pashnev}-\cite{Faux_Spector_kagan_Z}. We shall return to this point in Subsection \ref{Sec_Particular_case}.

In this work, some investigations of $\mathcal{N}=2$ supersymmetry with one (real) central charge is carried out in mechanics and two-dimensional field theory. We consider a similar superalgebra adopted in refs. \cite{Faux_Spector_Z,Faux_Spector_kagan_Z} and we propose an alternative implementation  of this supersymmetry through superspace approach. Here, we shall present a prescription to implement this extended supersymmetry by means of deformed covariant derivatives.  

The paper is organized as follows: in Section \ref{Sec_Mechanics}, we discuss the $\mathcal{N}=2$ supersymmetric mechanics with central charge.  We focus on the construction of the superspace formulation, central charge transformation and propose deformed  derivatives. In this context, we present a deformation of one-dimensional non-linear sigma model and revise the particular case discussed in ref. \cite{Faux_Spector_Z}. Then, a comparison between the two supersymmetric implementation is established. After that, in Section \ref{Sec_Field_Theory}, we turn our attention to supersymmetric field theories. Here, we develop a possible implementation of this superalgebra in two-dimensions, where a new coordinate is related to central charge. In this scenario, we present a model  with topological (bosonic) configurations and obtain a non-trivial fermionic solution. Finally, in Section \ref{Sec_Conc_Rem}, we display our Concluding Comments.

\section{Supersymmetric Mechanics with Central Charge} \label{Sec_Mechanics}
\indent 

In  $\mathcal{N}=2$ supersymmetric  one-dimensional systems (mechanical case), we have two (real) supercharges, $Q_1$ and $Q_2$, which satisfy the following superalgebra with the Hamiltonian $(H)$:  $\, Q^{2}_{1}=Q^{2}_{2} = H$ and $\left[ Q_1 , H \right] = \left[ Q_2 , H \right] = 0$.  In this work, we are going to deal with a possible extension of this case by including one real central charge $(Z)$ in the anti-commutator  $\lbrace Q_{1},Q_{2}\rbrace = 2Z$. We shall adopt a complex notation for the supercharges, namely, $ Q=\frac{1}{\sqrt{2}}\left(Q_{1}+iQ_{2}\right) \,$ and $ \bar{Q}=\frac{1}{\sqrt{2}}\left(Q_{1}-iQ_{2}\right) $, which leads to 
\begin{eqnarray}
\lbrace Q,\bar{Q} \rbrace = 2H \, , \label{alg_susy_1} \\
\left[ Q,H \right] = \left[ \bar{Q},H \right] = 0  \, , \label{alg_susy_2} \\
Q^2 = i Z \, \; , \, \; \bar{Q}^2 = -i Z \, . \label{alg_susy_3} 
\end{eqnarray}

With the aforementioned relations, one can check that $\left[ Q, Z \right] = \left[ \bar{Q} , Z \right] = \left[ H , Z \right] = 0$, i.e., $Z$ has the properties of a central charge. We also highlight that a similar superalgebra was considered in refs. \cite{Faux_Spector_Z,Faux_Spector_kagan_Z}, with only different conventions in eq. \eqref{alg_susy_3}.

In order to implement this extended supersymmetry through a superfield approach, we first introduce two (complex) Grassmann parameters, $\theta$ and $\bar{\theta}$, such that the superspace can be described by  $(t; \theta, \bar{\theta})$, where $t$ is the time parameter. Throughout this work, the Grassmann derivatives are understood in the sense of acting from left to the right,  e.g., $ \frac{\partial}{ \partial \theta} \left( \bar{\theta} \theta \right) = - \bar{\theta} $. 
Furthermore, we draw the attention to the procedure we adopt to realize the representation of central charge. Our proposal consists in implementing the central charge by means of a deformation of the supersymmetric generators. Now, to read the explicit form of this deformation, we act with the supersymmetric transformation on a  supermultiplet and, by imposing a number of conditions  that we are going later on to present, we get the final form of the deformation.

The superalgebra above can be realized in a differential representation. To achieve this, we define $\delta^H = i \partial_t $ and the following (deformed) supercharge operators: 
\begin{eqnarray}
\delta^{Q}&=&\partial_{\theta}+i\bar{\theta}\partial_{t}+i\theta\delta^{z} \, , \label{deformed_Q_1} \\
\delta^{\bar{Q}}&=&\partial_{\bar\theta}+i\theta\partial_{t}-i\bar{\theta}\delta^{z} \, ,\label{deformed_Q_2}
\end{eqnarray}
where we have used the notation $\partial_t = \partial/\partial t$, $\partial_\theta = \partial/\partial \theta$ and  
$\partial_{\bar{\theta}} = \partial/\partial \bar{\theta}$. 

Moreover, we  point out an important comment: it is worthy to introduce a deformation $\delta^z$ in the (supersymmetric) covariant derivatives.  Similarly to the case of the supercharges, eqs. \eqref{deformed_Q_1} and \eqref{deformed_Q_2}, we  define 
\begin{eqnarray}
\mathcal{\mathfrak{D}} &=& \partial_{\theta}-i\bar{\theta}\partial_{t}-i\theta\delta^{z} \, , \label{deformed_dev_1} \\
\bar{\mathcal{\mathfrak{D}}} &=& \partial_{\bar\theta}-i\theta\partial_{t}+i\bar{\theta}\delta^{z} \, ,
\label{deformed_dev_2}
\end{eqnarray}
which satisfy $ \lbrace \mathcal{\mathfrak{D}} , \bar{\mathcal{\mathfrak{D}}}  \rbrace = -2i \partial_t $ and have vanishing anti-commutation relations with the supercharges, $  \lbrace\delta^{Q}, \mathcal{\mathfrak{D}} \rbrace = \lbrace\delta^{Q}, \bar{\mathcal{\mathfrak{D}}} \rbrace = \lbrace\delta^{\bar{Q}}, \mathcal{\mathfrak{D}} \rbrace = \lbrace\delta^{\bar{Q}},\bar{\mathcal{\mathfrak{D}}} \rbrace = 0  \, . $

By using these (deformed) covariant derivatives instead of the usual $D=\partial_{\theta}-i\bar{\theta}\partial_{t}$ and  $\bar{D} = \partial_{\bar\theta}-i\theta\partial_{t}$, we automatically take into account the contribution of central charge and assure the extended supersymmetric invariance of the action formulated in terms of the covariant derivatives and superfields. For a trivial central charge transformation, we recover the usual $\mathcal{N}=2$ supersymmetry. On the other hand, one could use the derivatives $D$ and $\bar{D}$, as done in refs. \cite{Faux_Spector_Z,Faux_Spector_kagan_Z}. However, in this situation, one should perform a carefully analysis of the supersymmetric transformation of the Lagrangian (in components) and add some counter-terms  in order to maintain the invariance. Here, we claim that  these additional terms are exactly the ones generated by the derivatives $\mathcal{\mathfrak{D}}$ and $\bar{\mathcal{\mathfrak{D}}}$.
We shall return to this point in more details in the  subsection \ref{Sec_Particular_case}, where a particular case will be analyzed. 

At this moment, it is advisable to point out that, in this Section, the introduction of central charge is not associated with an extra coordinate in  superspace. In other words, the superspace is parametrized by $(t, \theta, \bar{\theta})$ and the  central charge is implemented through deformation of supercharges and covariant derivatives. 
 
The introduction of supercharges and deformed derivatives in connection with central charge is not exclusive to this work. For instance, in ref. \cite{Ivanov_Sidorov_PRD}, the authors investigated a central charge $(\Sigma)$ in the relation $\left\{ Q , \bar{Q}\right\} = 2 \left( H - \Sigma \right)$ and the invariance of the action is also implemented in a more simple way by using the superfields and the correspondent deformed derivatives.

As already mentioned, one of the goals in this work is to study in what multiplet we can introduce a non-trivial central charge transformation related to the superalgebra \eqref{alg_susy_1}-\eqref{alg_susy_3}. Here, we consider the multiplet $(1,2,1)$, described by 
one bosonic, two fermionic (Grassmann) and one auxiliary bosonic coordinates. The correspondent superfield is given by
\begin{equation}
X \equiv X[t; \theta, \bar{\theta} \, ] = x(t) + i\theta \, \xi(t) + i\bar{\theta} \, \bar{\xi}(t) + \theta \bar{\theta} \, \mathcal{W}(t) \, .
\label{superfield}
\end{equation}

Let us first establish the supersymmetric transformation of these components. By using $ \delta = \epsilon \delta^{Q} +\bar{\epsilon}\delta^{\bar{Q}} $, with  complex Grassmann parameters $\epsilon$ and $\bar{\epsilon}$, one can show that 
\begin{eqnarray}
\delta x &=&  i\epsilon \, \xi + i\bar{\epsilon} \, \bar{\xi} \, , \label{x_transf} \nonumber\\
\delta\xi &=&  - \bar{\epsilon} \, \dot{x}-i\bar{\epsilon} \, \mathcal{W} -\epsilon \, \delta^{z}x \, ,\label{xi_transf}\nonumber\\
\delta\bar{\xi} &=& -\epsilon \, \dot{x}+ i \epsilon \, \mathcal{W} + \bar{\epsilon}\, \delta^{z}x \, ,\label{bar_xi_transf}\nonumber \\
\delta\mathcal{W} &=& \frac{d}{dt} \left( \epsilon \,\xi - \bar{\epsilon} \, \bar{\xi} \, \right)
- \epsilon \, \delta^{z}\bar{\xi} -\bar{\epsilon} \, \delta^{z}\xi \, . \label{W_transf}
\end{eqnarray}

Now we can proceed to fix the central charge transformation. In the case of real superfield, we note that  $ \delta \xi $ is the complex conjugation of $\delta \bar{\xi}$, thereby $\delta^z x$ must be bosonic and pure imaginary. Moreover,   $\delta \mathcal{W}$ is bosonic, then $\delta^z \xi $ and $\delta^z \bar{\xi}$ should be fermionic. At this point, we suggest that $\delta^z x = i \mu$, where  $\mu $ is a constant (real) parameter. Bearing this in mind and applying the relation $\left[ \delta^z , \delta \right] = 0$ in all superfield components, one can arrive at the conditions $\delta^z \xi = 0 =\delta^z \bar{\xi}$ and $\delta^z \mathcal{W} = 0$. Actually, the main reason to fix this particular transformation is that $\delta \mathcal{W} = \frac{d}{dt} (...)$, which guarantees the invariance of the action in the superfield approach. For example, if we do not consider a constant, namely, $\delta^z x = i f(x)$, with $f(x)$ being an arbitrary function of $x$, we do not obtain a total derivative in $\delta\mathcal{W} $.

Finally, we remember that in  $\mathcal{N}=2$ supersymmetry (without $Z$), the quiral $(\phi)$ and anti-quiral $(\bar{\phi})$ superfields have been defined in a manner that satisfy the conditions $\bar{D} \phi = 0$ and $D \bar{\phi} = 0$. By using a K\"{a}hler (pre)potential, $K(\phi, \bar{\phi})$, these superfields were applied in the study of supersymmetric models associated with eletromagnetic interaction of a point particle \cite{Clark_Love_Nowling,Helayel_3}. Here, we emphasize that the imposition of the aforementioned conditions with the deformed derivatives leads to a trivial central charge transformations.
  

\subsection{Deformed non-linear sigma model} \label{Sec_Deformed_case}
\indent

The non-linear sigma models in connection with extended supersymmetric mechanics have been a subject of intense investigation. In refs. \cite{Coles_Papadopoulos}-\cite{Delduc_Ivanov}, the authors discussed  some cases with $\mathcal{N} > 2$ supersymmetries. The inclusion of central charges was done in \cite{Faux_Spector_Z,Faux_Spector_kagan_Z} and \cite{Bellucci_Nersessian} for $\mathcal{N}=2$ and $\mathcal{N}=4$, respectively. We also highlight some deformations involving $2-$ and $4-$forms, which  are related to torsion-like contributions \cite{Fedoruk_Ivanov_Smilga}  and  applications to Black-Hole \cite{Gibbons_Papadopoulos_Stelle}. For a review of one-dimensional non-linear sigma model and some aspects of geometry and topology, we point out  the works \cite{Hull} \cite{Wasay}. In the study of super-particles, one can introduce the so-called  tensorial central charges (see \cite{Bandos_Lukierski}$-$\cite{Kuznetsova_Toppan_2005} and  references therein), which are responsible to fix some constraints in the equations of motion and have found applications in higher-spin models.

Having established the superspace formulation,  let us investigate a possible application. Here, we propose what we refer to as our deformed non-linear sigma-model in terms of the deformed derivatives, a multiplet of real superfields $X^a$ and an arbitrary metric $g_{ab}(x)$. Then, the supersymmetric model is described by the following action  
\begin{eqnarray}
S & = & \int dtd\bar{\theta}d\theta \left[ \frac{1}{2}g_{ab}(X)\mathfrak{D}X^{a}\bar{\mathfrak{D}}X^{b}\right] \, , \label{full_action} \end{eqnarray}
%
which can be written in components as
\begin{eqnarray}
S & = & \int dt\,\bigg\{\frac{g_{ab}}{2}\bigg[\dot{x}^{a}\dot{x}^{b}+\mathcal{W}{}^{a}\mathcal{W}{}^{b}+i\xi^{a}\dot{\bar{\xi}}^{b}-i\dot{\xi}^{a}\bar{\xi}^{b}\nonumber \\
 & + & i\xi^{a}\delta^{z}\xi^{b}+i(\delta^{z}\bar{\xi}^{a})\bar{\xi}^{b}+\bigg(\delta^{z}x^{a}\bigg)\bigg(\delta^{z}x^{b}\bigg)\bigg]\nonumber \\
 & - & \frac{i\partial_{j}g_{ab}}{2}\bigg[\bigg( \dot{x}^{b} \bar{\xi}^{j}\xi^{a} -i \mathcal{W}^{b} \bar{\xi}^{j}\xi^{a} -\bar{\xi}^{j}\bar{\xi}^{b}\delta^{z}x^{a}\bigg)\nonumber \\
 & + & \bigg( \dot{x}^{a}\xi^{j}\bar{\xi}^{b}+i\mathcal{W}^{a}\xi^{j}\bar{\xi}^{b} + \xi^{j}\xi^{a}\delta^{z}x^{b} \bigg) \nonumber \\
 & - & i\mathcal{W}{}^{j}\xi^{a}\bar{\xi}^{b}\bigg]\ -\frac{1}{2} \left( \partial_{j}\partial_{k}g_{ab} \right) \, \xi^{j}\bar{\xi}^{k}\xi^{a}\bar{\xi}^{b}\bigg\}\, .
\label{action_11} \end{eqnarray}

By taking $\delta^z x^a = i \mu^a$ and $\delta^z \xi^a = 0$, we obtain two new contributions: one coupled to the metric and other with its first derivative.  Notice that for an euclidean metric $(g_{ab} = \delta_{ab})$,  we obtain a trivial result, namely, the Lagrangian (without central charge) is shifted by a constant. Therefore, only in specific curved spaces, one can introduce a non-trivial deformation.

\subsection{A particular Non-linear Sigma-model} \label{Sec_Particular_case}
\indent
 
In this subsection, let us revise a particular one-dimensional non-linear sigma model described in ref. \cite{Faux_Spector_Z} under a distinct point of view. The model  consists of a point-particle restrict to the cylinder-like topology with a variable radius. More specifically, we consider a particular metric  $ds^2 = g_{ab}(x) d x^a \, dx^b \, \; (a,b=1,2) \,$, where $g_{ab} = \textrm{diag}(1, h(x^{1}))$ and $h(x^1)$ denotes an arbitrary non-negative function. Hence, $x^1$ is the axial coordinate with $\sqrt{h(x^1)}$ being the cylinder radius and $x^2$ corresponds to the angular coordinate.   

In the supersymmetric description of this model, we have two real superfields $X^1$ and $X^2$ with the following central charge transformations: $\delta^Z x^1 = i \mu$ and $\delta^Z x^2 = 0$.
%
%
%
After considering  these particular metric and transformations in action \eqref{action_11}, 
one arrives at the correspondent Lagrangian
\begin{eqnarray}
L & = & \frac{1}{2}\,\bigg[(\dot{x}^{1})^{2}+i\bigg(\xi^{1}\,\dot{\bar{\xi}}^{1}-\dot{\xi}^{1}\,\bar{\xi}^{1}\bigg)+(\mathcal{W}^{1})^{2}\bigg]\nonumber \\
 & + & \frac{1}{2}\,h\,\bigg[(\dot{x}^{2})^{2}+i\bigg(\xi^{2}\,\dot{\bar{\xi}}^{2}-\dot{\xi}^{2}\,\bar{\xi}^{2}\bigg)+(\mathcal{W}^{2})^{2}\bigg]\nonumber \\
 & + & \frac{1}{2}\,h'\,\bigg[-i\dot{x}^{2}\bigg(\xi^{1}\,\bar{\xi}^{2}+\bar{\xi}^{1}\,\xi^{2}\bigg)+\mathcal{W}^{2}\bigg(\xi^{1}\,\bar{\xi}^{2}-\bar{\xi}^{1}\xi^{2}\bigg)\nonumber \\
 & - & \mathcal{W}^{1}\,\xi^{2}\bar{\xi}^{2}\bigg]-\frac{1}{2}\,h''\,\xi^{1}\,\bar{\xi}^{1}\,\xi^{2}\,\bar{\xi}^{2}\nonumber \\
 & + & \frac{1}{2}\,\mu\,h'\,\bigg(\xi^{1}\,\xi^{2}-\bar{\xi}^{1}\,\bar{\xi}^{2}\bigg)-\frac{1}{2}\,\mu^{2}\,h\,,
\label{11} \end{eqnarray}
where $h'$ and $h''$ denote the first and second derivatives of $h(x^1)$, respectively. 

According to ref. \cite{Faux_Spector_Z}, after carrying out the canonical quantization and imposing the superalgebra on the operators, it is possible to recognize a connection involving the central charge parameter $(\mu)$ and the quantum number $(\nu)$ of the angular momentum, given by the map duality $\mu \leftrightarrow \nu$. In addition, this duality also maps the radius $R =\sqrt{h(x^1)}$ and $x^1$ with the following simultaneous transformations: $R \leftrightarrow 1/R$ and $x^1 \leftrightarrow - x^1$.

Finally, we notice that the Lagrangian \eqref{11} is equivalent to the one obtained in ref. \cite{Faux_Spector_Z}. It only differs from some signs and $i-$factors which are due our different conventions. At this point, it is interesting to compare both methodologies.
In ref. \cite{Faux_Spector_Z}, the authors considered the 
following action
\begin{eqnarray}
S_0 &=& \int dtd\bar{\theta}d\theta \, \bigg[\frac{1}{2} \,  g_{ab}(X) \, DX^{a}\bar{D}X^{b}\bigg] \, ,
\label{acao_S_0} \end{eqnarray}
where $D=\partial_{\theta}-i\bar{\theta}\partial_{t}$ and  $\bar{D} = \partial_{\bar\theta}-i\theta\partial_{t}$ correspond  to  derivatives without central charge. Then, eq. \eqref{acao_S_0} is decomposed in components and a  analysis of supersymmetric transformation (with central charge) leads to a non-invariant action. In order to restore the   invariance, some corrections terms  $(c-t)$ are needed, i.e., a new contribution $S_{c-t}$ is added to $S_0$ such that the complete action $S = S_0 + S_{c-t}$ remains invariant. Here, we claim that this complete action coincides with eq. \eqref{full_action}. Therefore, the deformed derivatives automatically includes the correspondent $L_{c-t}$, namely, the last two terms in eq. \eqref{11} with  parameters $\mu$ and $\mu^2$.

\section{Supersymmetric Field Theory with Central Charge} \label{Sec_Field_Theory}
\indent

In this section, we present another point of view to investigate this extended supersymmetry. Our proposal is to introduce a new coordinate $v$ related to central charge $Z$. 
It is worthy to comment that the introduction of extra bosonic coordinates $-$ and, consequently, an extended superspace and deformed derivatives $-$ is not exclusive to this work. We highlight some works \cite{Sohnius_1978}$-$\cite{Lukierski_Rytel_82} in which this approach has been adopted in other contexts. For instance, in ref.  \cite{Azcarraga_Lukierski_86}, the authors developed a $N-$extended Poincaré superalgebra with $\frac{N(N-1)}{2}$ tensorial central charges and applied this formulation in higher-dimensional super-particle.

Here, we shall generalize the  procedure described in ref. \cite{Clark_Love_Nowling} for $\mathcal{N}=2$ supersymmetric mechanics (without $Z$). Initially, let us define a group element 
\begin{equation}  G(t, v , \theta , \bar{\theta }) = e^{itH + i v Z + i\theta Q + i 
\bar{Q} \bar{\theta } } \, . \label{SUSY_Z_v_0} \end{equation}

By using the superalgebra, eqs. \eqref{alg_susy_1}-\eqref{alg_susy_3}, and the well-known Baker-Campbell-Hausdorff identity, $ e^A e^B = e^{A+B} \, e^{(1/2) \left[A,B \right]} \, ,$  where $ \left[[A,B],A \right] = 0 = [[A,B],B] \, ,$
one may verify the group structure  with the   multiplicative rule
\begin{eqnarray*}
G(t',v',\theta',\bar{\theta}')\,G(t,v,\theta,\bar{\theta}) & =
\end{eqnarray*}
\begin{eqnarray}
G\left(t+t'+i(\theta'\bar{\theta}+\bar{\theta}'\theta),v+v'+(\theta'\theta-\bar{\theta}'\bar{\theta}),\theta+\theta',\bar{\theta}+\bar{\theta}'\right),\nonumber \\
\label{SUSY_Z_v_1}
\end{eqnarray}
which leads to the following translation in the new superspace
\begin{eqnarray*}
(t,v,\theta,\bar{\theta})\rightarrow
\end{eqnarray*}
\begin{eqnarray}
\left(t+t'+i(\theta'\bar{\theta}+\bar{\theta}'\theta)\,,v+v'+(\theta'\theta-\bar{\theta}'\bar{\theta})\,,\,\theta+\theta'\,,\,\bar{\theta}+\bar{\theta}'\right)\,.\nonumber\\
\end{eqnarray}
Notice that a real translation acts on $v$ in a similar way with  time, namely, in both cases appear bilinear of the  Grassmann parameters $(\theta, \bar{\theta})$.

Starting with the group element $G(t,v,\theta, \bar{\theta})$ described previously, we establish the superfield as 
\begin{equation} X (t,v,\theta, \bar{\theta}) = G(t,v,\theta, \bar{\theta}) \, X(0,0,0,0) 
\, G^{-1} (t,v,\theta, \bar{\theta}) \, . \end{equation}

Now we consider an infinitesimal transformation and obtain the differential representation to  supercharges and central charge. By using infinitesimal parameters  $ (t',v', \theta', \bar{\theta}') \rightarrow (\varepsilon , \zeta , \epsilon , \bar{\epsilon})$ and the multiplicative rule \eqref{SUSY_Z_v_1}, we have 
$$ G (\varepsilon , \zeta , \epsilon , \bar{\epsilon}) \, X(t,v, \theta, 
\bar{\theta}) \, G^{-1}(\varepsilon , \zeta , \epsilon , \bar{\epsilon}) = $$
\begin{equation} = X \left( t + \varepsilon + i ( \epsilon \, \bar{\theta} + 
\bar{\epsilon} \,
\theta ), v + \zeta + ( \epsilon \, \theta - \bar{\epsilon} \, \bar{\theta}) \, , 
\, \theta + \epsilon , \bar{\theta} + \bar{\epsilon}  \right) \, .
\label{SUSY_Z_v_2} \end{equation}

A general operator $\mathcal{O}$ satisfies $ i [ \mathcal{O} , X] \sim \delta^{\mathcal{O}}X$. Thus, from  eq. \eqref{SUSY_Z_v_2}, one may expand both sides and obtain the following differential representation:   
\begin{eqnarray} \delta^{H} &=& i \partial_t \, \, , \, \, \delta^{Z} = - i \partial_v \, , \label{SUSY_Z_v_3} \\ \delta^{Q} &=& \partial_\theta + i \bar{\theta} \partial_t + \theta  \partial_v \, \, , \, \, \delta^{\bar{Q}} = \partial_{\bar{\theta}} + i \theta \partial_t -  \bar{\theta} \partial_v \, . \label{SUSY_Z_v_4}\end{eqnarray}
%
%
%

At this point, we realize that the central charge behaves like a  momentum operator of the ``extra-dimension" $v$. In comparison with the supercharges in mechanical case, eqs.
\eqref{deformed_Q_1} and \eqref{deformed_Q_2}, we have an analogous structure, but now  central charge is completely fixed, $\delta^Z = - i \partial_v$, independently of the superfield components. These differential operators satisfy the same superalgebra: $ \left\{ \delta^{Q} , \delta^{\bar{Q}} \right\} = 2 \,
\delta^{H}$, $(\delta^Q)^2 = i \delta^Z $ and $ (\delta^{\bar{Q}})^2 = -i \delta^Z$, with $\delta^{H} $ and $\delta^{Z}$ commuting with all operators.

In order to obtain the covariant derivatives, we consider an alternative multiplication to the right of the group elements, $ G(t, v , \theta , \bar{ \theta} ) \, G(t', v' , \theta' , \bar{\theta}' )   $, which leads to
\begin{equation} \mathfrak{D} = \partial_\theta - i \bar{\theta} \partial_t - \theta \partial_v 
\, \, , \, \, 
\bar{\mathfrak{D}} = \partial_{\bar{\theta}} - i \theta \partial_t +  \bar{\theta} 
\partial_v \, . \label{SUSY_Z_v_5} \end{equation}

These covariant derivatives exhibit the same representation of the deformed derivatives in the mechanical case (now with $\delta^Z = - i \partial_v$) and anti-commutation rules with the supercharges $\delta^{Q}$ and $\delta^{\bar{Q}}$. In both mechanical and field theory situations, the main role of the deformed derivatives is to provide a description with a manifested (extended) supersymmetry. In other words, by using the superfields and deformed derivatives, the invariance of the action is guaranteed. Moreover, these deformed derivatives are intrinsically related to the central charge by the following algebra: $ \bar{\mathfrak{D}}^2 = i \delta^Z$ and $\mathfrak{D}^2 = -i \delta^Z$. Thus, for a vanishing central charge, the derivatives shall be nilpotent.  

Having established the supercharges and covariant (deformed) derivatives, we turn our attention to the discussion of superfield and  supersymmetric transformation. We define a real (bosonic) superfield  as
\begin{equation} X = f_1(t,v) + i \theta \psi(t,v) + i 
\bar{\theta} 
\bar{\psi}(t,v) + f_2(t,v) \, \theta \bar{\theta} \ . \label{SUSY_Z_v_6} \end{equation} 

It is important to mention that, in this formulation, we do not have  a classical mechanics description, because now we deal with the component fields $f_1, f_2 , \psi $ and $\bar{\psi}$, which in general depend on $(t,v)$. However, by taking a dimensional reduction, one can arrive at mechanical case. For example, in the trivial reduction, $\partial_v \,(\textrm{all fields}) = 0$, we recover the usual $\mathcal{N} =2$ supersymmetry (without central charge) with the identification $( f_1 , \psi , \bar{\psi} , f_2) \, \rightarrow \, (x, \xi, \bar{\xi}, \mathcal{W})$.
  
Let us obtain the component transformations with this kind of supersymmetry. 
By taking the Taylor expansion of the right-hand side of
eq. $(\ref{SUSY_Z_v_2})$ and comparing with $\delta X \equiv 
 \delta f_1 + i \theta \, \delta \psi + i \bar{\theta} \, \delta \bar{\psi} + \theta 
\bar{\theta}\, \delta f_2$,
we have the following variations
\begin{equation} \delta f_1 = i \, \epsilon \, \psi + i \, \bar{\epsilon} \, 
\bar{\psi} + \varepsilon \dot{f}_1 + \zeta \, \partial_v f_1 
\, , \label{SUSY_Z_v_7} \end{equation}
\begin{equation} \delta \psi =  i \epsilon \, \partial_v f_1 - i \, \bar{\epsilon} 
\, f_2 - \bar{\epsilon} \, \dot{f}_1 + \varepsilon \dot{\psi} + \zeta \, \partial_v \psi
\, , \label{SUSY_Z_v_8} \end{equation}
\begin{equation} \delta f_2 = \frac{\partial}{ \partial t} \left( \epsilon \, \psi - \bar{\epsilon} 
\,\bar{\psi} + \varepsilon f_2  \right) + \frac{\partial}{ \partial v} \left( i \epsilon \, \bar{\psi} 
+ i \bar{\epsilon} \, \psi + \zeta f_2 \right) \, . \label{SUSY_Z_v_9} \end{equation}

Once $f_2$ transforms as partial derivatives of $v$ and $t$, one may describe an invariant action  in terms of real superfield and covariant derivatives. For supersymmetric transformations, we just need to fix $\zeta = \varepsilon = 0$ in the  last variations, such that $ \delta^{SUSY} = \epsilon \, \delta^{Q} + \bar{\epsilon} \, 
\delta^{\bar{Q}}\, , $
where $\delta^{Q}$ and $\delta^{\bar{Q}}$ are given by eq.$(\ref{SUSY_Z_v_4})$. In this case, the component transformations are very similar to the mechanical case, eqs. \eqref{W_transf}.

Finally, we emphasize that in our formalism the chiral and anti-chiral superfields do not depend on $v$ and have the same structure of the supersymmetric model without central charge. 


\subsection{Topological configurations in (1+1)D}

In this subsection, we discuss an application of the previous formalism. We investigate a particular model in two-dimensions. Using  the covariant derivatives $(\ref{SUSY_Z_v_5})$ and  real superfield $(\ref{SUSY_Z_v_6})$, we propose the following action 
\begin{equation} S = \int dt \, dv \, d\bar{\theta} \, d\theta  \, \left[  
\frac{1}{2} \, \mathfrak{D} X \, \bar{\mathfrak{D}} X + U(X) \, \right] \, , 
\label{kink_1} \end{equation}
where $U(X)$ denotes an arbitrary superpotential.

This action leads to the following Lagrangian density
$$ \mathcal{L} = \frac{1}{2} \, (\dot{f_1})^2 - \frac{1}{2} \, (\partial_v f_1)^2 + \frac{(f_2)^2}{2} 
+ f_2 \, \frac{\partial U}{\partial f_1} \,  $$
\begin{equation} + \psi \bar{\psi} \, \frac{\partial^2 U}{\partial f_1^2} + \frac{i}{2} 
\, (\psi \dot{\bar{\psi}} - \dot{\psi} \bar{\psi} ) + \frac{1}{2}  \left( \,
(\partial_v \bar{\psi}) \bar{\psi} + \psi (\partial_v \psi) \, 
\right) \, . \label{kink_2} \end{equation}

Hence, we obtain a Klein-Gordon profile (first two terms), so we again note that $v$ and $Z$ can be interpreted as spatial coordinate and momentum operator $(\delta^{Z} = - i \, \partial_v)$, respectively. In this case, we concluded that the role of central charge is to accommodate  field theory in $(1+1)D$. 

Since the auxiliary field $f_2(t,v)$ does not have dynamics, it can be eliminated by its equation of motion, 
$f_2(t,v) = - \partial U / \partial f_1 $, resulting on the on-shell Lagrangian density
$$ \mathcal{L}_{\textrm{on-shell}} = \frac{1}{2} \, (\dot{f}_1)^2 - \frac{1}{2} \, (\partial_v f_1)^2  
- \frac{1}{2} \left( \frac{\partial U}{\partial f_1} \right)^2 
+ \psi \bar{\psi} \, \frac{\partial^2 U}{\partial f_1^2} \, + $$
\begin{equation} + \frac{i}{2} \, 
(\psi \dot{\bar{\psi}} - \dot{\psi} \bar{\psi} ) + \frac{1}{2}  \left( \,
(\partial_v \bar{\psi}) \bar{\psi} + \psi (\partial_v \psi) \, 
\right) \, . \label{kink_4} \end{equation}

The equations of motion are given by 
\begin{equation} \ddot{f}_1 - \partial^2_v f_1 + U' \, U'' - \psi \bar{\psi} \, U''' = 0 
\, , \label{kink_5} \end{equation}
\begin{equation} i \, \dot{\psi} - \, \partial_v \bar{\psi} - \psi \, U'' = 0 \,  , \label{kink_6}  \end{equation}
where  we used $U' \equiv \partial U / \partial f_1 $ and similarly for higher derivatives. 

Here, we  notice that  these equations of motion can accommodate topological configurations. Let us initially consider a trivial fermionic sector, $\psi (t,v) = \psi^{(1)} (t,v) = 0$. Then, for a particular case  $f_1(t,v) = f_1^{(1)}(v)$,  satisfying 
\begin{equation} \partial_v f_1^{(1)} = \pm U'( f_1^{(1)} ) \, ,
\label{kink_7} \end{equation}
we obtain a possible solution. This first order equation can describe some topological configurations. We only need to specify the potential $U(f_1)$. Before do that, we would like to point out some comments about the fermionic sector.

It is interesting to highlight that we can take advantage of the supersymmetric structure to get a set of solutions with non-trivial fermionic sector. We shall adopt a similar procedure discussed in ref. \cite{Susy_oscillons}, where oscillinos (fermionic) solutions were obtained by means of supersymmetric transformations on the oscillons (bosonic configurations). 

First, one may verify that, by applying supersymmetric transformations, eqs. \eqref{kink_5} and \eqref{kink_6} are mapped into each other. That is exactly the sense of supersymmetry. In order to obtain a non-trivial fermionic solution with topological structure in the bosonic sector, we use the on-shell supersymmetric transformations, namely, eqs. 
$(\ref{SUSY_Z_v_7})$ and  $(\ref{SUSY_Z_v_8})$  with $ \varepsilon = \zeta = 0 $ and $f_2= - U' $. In other words, we need to consider a supersymmetric  perturbation of the previous solution,  
\begin{equation} f_1^{(2)} = f_1^{(1)} + \delta f_1^{(1)} \equiv f_1^{(1)} \, , 
\end{equation}
\begin{equation} \psi^{(2)}  = \psi^{(1)} + \delta \psi^{(1)} \equiv \delta \psi^{(1)} = 
i \epsilon \, \partial_v f^{(1)}_1 + i \bar{\epsilon} \, U'  \, , \label{pert_psi} \end{equation}
which are also solutions of the equations of motion $(\ref{kink_5})$ and $(\ref{kink_6})$.

Finally,  let us present a particular case. If we fix the arbitrary function $U(f_1)$ such that 
\begin{equation} \frac{1}{2} \, \left[ U'(f_1 ) \right]^2 =  A \,   \left[ 1 - \cos \left( \frac{2 \pi}{F} \, f_1  \right)  \right]
 \, , \label{kink_8} \end{equation} 
with $A$ and $F$ being constant parameters, then we obtain a new supersymmetric version of the sine-Gordon model \cite{Notes_Hooft}. In this case, the bosonic solution is given by
\begin{equation}
f_1^{(2)}(v) = \frac{2 F}{\pi} \, \arctan \left[ e^{ m (v - v_0) } \right] \, ,
\end{equation}
where $v_0$ is an arbitrary constant and $m \equiv 2\pi \sqrt{A}/F$.

By using the supersymmetric pertubation, eq. \eqref{pert_psi}, one can arrive at the following fermionic (static) solution
\begin{eqnarray}
\psi^{(2)}(v) & = & i\,\epsilon\,4\sqrt{A}\,\frac{e^{m(v-v_{0})}}{1+e^{2m(v-v_{0})}}\nonumber \\
 & \pm & i\,\bar{\epsilon}\,\left\{ 2A\left[1-\cos\left(\frac{2\pi}{F}f_{1}^{(2)}(v)\right)\right]\right\} ^{1/2}\,.
\end{eqnarray}

Finally, it is worthy to comment that, by construction, this fermionic solution has a trivial condensate, $ \bar{\psi}^{(2)} \, \psi^{(2)} = 0$. This could indicate a possible relation between the extended superalgebra (with central charge and coordinate $v$) and other two-dimensional supersymmetries. We shall return to this point in our  conclusions.

\section{Concluding Comments} \label{Sec_Conc_Rem}
\indent 

Initially, we have discussed the $\mathcal{N}=2$ supersymmetric mechanics with one (real) central charge for the multiplet $(1,2,1)$. A prescription to obtain deformed $\mathcal{N}=2$ models by central charge was developed. To establish this in a superfield approach, we have introduced deformed covariant derivatives, eqs. \eqref{deformed_dev_1} and \eqref{deformed_dev_2}, which take into account the new terms related to the central charge. As an  
application, we have obtained a deformation of one-dimensional nonlinear sigma-model. Also, we have  recast the  particular non-linear sigma-model of the ref. \cite{Faux_Spector_Z}  and shown an equivalence between the two prescriptions for the specific transformations given by eqs. \eqref{W_transf}.  
However, we have noticed that an introduction of deformed derivatives allows us to implement this extended supersymmetry in a simplest way,  once we maintain the superfields and it is not necessary to decompose the Lagrangian in components and add counter-terms to recover the supersymmetry. Bearing this in mind, we would like to point out some possible subjects of investigations. In the context of general non-linear sigma-model (with Riemann curvature), one may add the torsion and generalized torsion terms \cite{Fedoruk_Ivanov_Smilga}, namely,  some couplings involving $2-$ and $4-$form with $DX^i DX^j$, $DX^i DX^j DX^k DX^l$ and its complex conjugations. A deformation of these models  could be obtained through the following prescription proposed here: $D \rightarrow \mathcal{\mathfrak{D}}  $ and $\bar{D} \rightarrow \bar{\mathcal{\mathfrak{D}}}$. It would be interesting to investigate the quantization of these deformed models and possible new connections between central charge and symmetries (e.g., duality and mirror) or some restrictions to the manifold.

In the second part, we have considered an implementation of superalgebra \eqref{alg_susy_1}-\eqref{alg_susy_3}  in two-dimensional field theory. Among the lines discussed in the Introduction \ref{Sec_Intro},  a particular point of view has been adopted here, namely, we  have introduced a new coordinate $v$ and interpreted the central charge as a momentum operator. With these assumptions, the supersymmetric transformations  are fully fixed, given by eqs. \eqref{SUSY_Z_v_7}-\eqref{SUSY_Z_v_9}, where $\delta^Z = -i \partial_v$. This interpretation allowed us to obtain a non-trivial fermionic solution through supersymmetric tranformations  rather than solving directly the fermionic equation of motion. 

Finally, we point out some possible investigations related to this supersymmetry in two dimensions. First, in order to accommodate charged matter and gauge fields, one could analyze the introduction of other multiplets, such as complex bosonic (fermionic) scalars and vectors superfields. Moreover, the connection between this superalgebra and other two-dimensional (Poincaré) supersymmetries  remains as a subject of further investigation. In particular, one could study the usual supersymmetry in two dimensions and redefine or drop out some Lorentz (boost) generators. Remembering that in two dimensions, it is possible to have a Majorana-Weyl fermion and implement the heterotic $(p,q)-$supersymmetries. 
This perspective is based on the fact that the Majorana-Weyl condition in two dimensions implies one degree-of-freedom with a trivial condensate. In our context, we have also obtained a similar situation for the fermionic solution $\psi^{(2)}(v)$.

\section*{Data Availability}
\noindent 

No data were used to support this study.

\section*{Conflicts of Interest}
\noindent 

The authors declare that there is no conflict of interest regarding the publication of this paper.

\section*{Acknowledgements}
\noindent 
We would like to thank Prof. J.A. Helay\"{e}l-Neto for  useful comments and reading the manuscript. LPRO is supported by the  National Council for Scientific and Technological Development (CNPq/MCTIC) through the PCI-DB funds, grant no. $30.0644/2017-5$. RCT is grateful to the Coordination for the Improvement of Higher Education Personnel (CAPES) for financial support under the program Doutorado Sandu\'{i}che no Exterior (PDSE), grant no. $88881.188419/2018-01$.

\end{document}